\documentclass[prl,twocolumn,superscriptaddress,amsmath,amssymb,aps,nofootinbib,floatfix]{revtex4-2}
\usepackage[utf8]{inputenc}
\usepackage{amsmath,amsthm,amssymb,amsfonts,braket}
\usepackage{placeins}[verbose]
\usepackage{tabularx, longtable, multirow}
\usepackage{graphicx}
\usepackage{gensymb} 
\usepackage{dcolumn}
\usepackage{bm,dsfont}
\usepackage[english]{babel}
\usepackage{comment}
\usepackage[dvipsnames]{xcolor}
\usepackage{siunitx}
\usepackage[normalem]{ulem}
\usepackage[T1]{fontenc}
\usepackage{physics}
\usepackage{mathtools}

\newcommand{\mapsfrom}{\mathrel{\reflectbox{\ensuremath{\mapsto}}}}

\begin{document}
\title{Boundary-Sensitive Quantum Order on Non-Amenable Lattices}

\author{Xingzhi Wang}
\email{xingzhiwang2024@u.northwestern.edu}
\affiliation{Department of Physics, Boston College, Chestnut Hill, MA 02467, USA.}
\affiliation{Department of Physics, Washington University, St.
Louis, MO 63130, USA}
\author{Zohar Nussinov}
\email{zohar@wustl.edu}
\affiliation{Department of Physics, Washington University, St.
Louis, MO 63130, USA}
\author{Gerardo Ortiz}
\email{ortizg@iu.edu}
\affiliation{%
Department of Physics, Indiana University, Bloomington IN 47405, USA}%
\affiliation{Institute for Advanced Study, Princeton, NJ 08540, USA}

\date{\today} 

\begin{abstract}
We show that quantum systems on non-amenable lattices, whose boundaries constitute an extensive fraction of the system, host novel forms of bulk ordering with no Euclidean counterpart. Combining boundary-retaining iDMRG with an exact symmetry-fixing duality, we investigate $p<5$ state clock models on the coordination-3 infinite Cayley tree. We find that between the ordered and disordered phases lies an emergent intermediate phase with bulk order induced by an arbitrarily weak boundary field rather than by conventional spontaneous symmetry-breaking. This establishes boundary-sensitive order as a distinct quantum ordering mechanism defying Landau's spontaneous symmetry-breaking paradigm. The intermediate phase realizes a bulk-boundary correspondence in which the unpolarized boundary selects a specific ordered bulk state. 

\end{abstract}

\maketitle

{\it Introduction.--} Negative curvature fundamentally alters the interplay between bulk and boundary physics. Hyperbolic lattices provide natural discrete settings for boundary conformality and holographic bulk–boundary correspondence~\cite{HaPPY2015,Holography1,Holography2}, and host phenomena absent in Euclidean lattices, including non-Abelian band structures~\cite{HyperbolicBand}, boundary-controlled many-body behavior~\cite{Gotz2024}, and extensive boundary-dominated states~\cite{HyperbolicChern1,HuangHyperbolicPhotonic,PatinoHyperbolicMechanical}. These physical effects have now been realized in circuit-QED, topolectrical, photonic, and mechanical platforms~\cite{HyperbolicChern2,KollarHyperbolic,LenggenhagerCircuit,ChenHyperbolicMatter,HuangHyperbolicPhotonic,PatinoHyperbolicMechanical}. 

Underlying these phenomena is the defining property of non-amenability: the boundary remains an extensive fraction of the system. Strikingly, this geometry gives rise to phases of matter that defy the conventional Landau paradigm~\cite{WangNussinovOrtiz2025}. On the infinite Cayley tree, for example, the absence of loops renders the Ising free-energy density analytic at all temperatures, while boundary recursion and bulk-energy variations diagnoses a mean-field-like transition in the bulk~\cite{Eggarter1974,CayleyIsing,LyonsNonamenable,MullerHartmann,IoffeExtremality,BleherPurity,Hyperising}, and the bulk ordering occurs only in the presence of a symmetry-breaking boundary~\cite{WangNussinovOrtiz2025,CCWU1,CCWU2,YJ}. Classical systems on more general hyperbolic lattices likewise exhibit two-stage transitions, with boundary-induced bulk order emerging before the free-boundary state orders~\cite{WangNussinovOrtiz2025,CCWU1,CCWU2,YJ,CTMRGH1,Lee2016,CTMRGH4}.

The regime between these two transitions was recently identified as a boundary-sensitive phase~\cite{WangNussinovOrtiz2025}. Here, a bulk observable $A$ invariant under the global symmetry of the Hamiltonian takes different thermodynamic-limit values depending on whether a boundary symmetry-breaking perturbation 
$h_\partial$ is removed before or after the system volume $\mathcal{ V}$ is sent to infinity,
\begin{eqnarray}
    \lim_{h_\partial\rightarrow0^+}\lim_{\mathcal{V}\rightarrow \infty}\langle A\rangle_{h_\partial} \neq \lim_{\mathcal{V}\rightarrow \infty}\lim_{h_\partial\rightarrow0^+}\langle A\rangle_{h_\partial}\;.
\label{eq:BSP}
\end{eqnarray}
The right-hand side is the symmetric state; the left-hand side is the arbitrarily weak boundary field selected (boundary-biased) state. An intermediate regime emerges whenever the biased state exhibits bulk order while the symmetric state remains disordered,
thereby defining a distinct phase.

A central question is whether such emergent structure persists in the quantum realm. The quantum-to-classical (Feynman or Suzuki-Trotter) correspondence, viewed as an isometry between the bond algebras of transfer matrices~\cite{CON2011,NO2011}, provides no mapping to known instances of classical boundary sensitivity. Moreover, boundary-induced ordering poses a challenge for numerical studies of the thermodynamic ($\mathcal V \rightarrow \infty$) limit: bulk optimization methods, such as Uniform Matrix Product States ~\cite{Nagaj2008,Li2012,ZaunerStauber2018,Lunts2021} and cavity approaches~\cite{LaumannCavity,KrzakalaPathIntegral}, do not capture the extensive boundary required to establish Eq.~\eqref{eq:BSP}. Conversely, boundary-retaining methods such as infinite Density Matrix Renormalization Group (iDMRG)~\cite{White1992,Otsuka1996,Friedman1997,Kumar2012} are susceptible to machine-precision symmetry breaking across the extensive boundary, making Eq.~\eqref{eq:BSP} indistinguishable from genuine spontaneous symmetry breaking. A definitive resolution therefore requires treating the full boundary-retained quantum system under \textit{exact} symmetry control~\cite{WangNussinovOrtiz2025}.

In this Letter we establish the existence of boundary-sensitive quantum order. We determine the quantum phase diagrams of the transverse-field Ising ($p=2$) and three-state clock ($p=3$) models on the coordination-3 infinite Cayley tree. The four-state ($p=4$) clock model, in turn, maps exactly onto two copies of the $p=2$ model \cite{OCN2012} and thus shares its phase diagram. We combine bond-algebraic dualities~\cite{CON2011,OCN2012} with boundary-retaining iDMRG to perform simulations directly on the specific $\mathbb{Z}_p$ symmetry sector that contains the ground state~\cite{WangNussinovOrtiz2025}. For both models, we identify three distinct phases: disordered, intermediate, and ordered. In the intermediate phase, the bulk orders in the boundary-biased state while remaining disordered in the symmetric state, in agreement with Eq.~\eqref{eq:BSP}. Remarkably, the boundary itself remains unpolarized throughout this regime and orders only upon entering the ordered phase, with an anomalously slow onset. Magnetization profiles of the \(p=2\) Ising model reveal the underlying mechanism: in the disordered and ordered phases, boundary perturbations are absorbed within the first few layers, as in Euclidean systems, whereas in the intermediate phase they penetrate arbitrarily deep beneath and into the bulk, giving rise to boundary-sensitive order.



{\it The model.--} Consider the quantum $p$-clock model on a graph $\mathrm{G}(\mathrm{V},\mathrm{E})$ \cite{OCN2012},
\begin{equation}
    H_{p\text{-clock}}
    =
    -\frac{J}{2}\sum_{e\in \mathrm{E}} U_i U_j^\dagger
    -\frac{h}{2}\sum_{i\in \mathrm{V}}V_i
    +\mathrm{h.c.}\;,
\label{eq:pclock}
\end{equation}
with $J,h>0$ and integer $p\geq 2$, and $e=(i,j)$. Here $U_i$ and $V_i$ are $p$-dimensional clock and shift unitary operators on site $i$ satisfying
\begin{align}
   V_i^p&= U_i^p= \mathbb{I} ;,\label{eq:clockalgebra1}\\
    V_i U_j&=
    \omega^{\delta_{ij}} U_j V_i \;,\label{eq:clockalgebra2}\\
   {[}V_i,V_j]&=[U_i,U_j]=0\;,
\label{eq:clockalgebra3}
\end{align}
where $\omega=\exp(\mathrm{i} 2\pi/p)$. The Hamiltonian has a global $\mathbb{Z}_p$ symmetry represented by the unitary operator
\begin{equation}
    \mathcal{U}
    =
    \prod_{i\in \mathrm{V}} V_i .
\label{eq:zp}
\end{equation}

For $p=2$, Eq.~\eqref{eq:pclock} reduces to the transverse-field Ising model,
\begin{equation}
    H_{\mathrm{Ising}}
    =
    -J\sum_{e\in \mathrm{E}}\sigma_i^z\sigma_j^z
    -h\sum_{i\in \mathrm{V}}\sigma_i^x .
\label{eq:Ising}
\end{equation}
For \(p=4\), Eq.~\eqref{eq:pclock} decomposes exactly into two identical copies of the Ising model, Eq.~\eqref{eq:Ising}, under the mapping~\cite{OCN2012}
\begin{align}
    U_i &= \frac{1}{2}\left[(1-\mathrm{i})\sigma_i^z+(1+\mathrm{i})\tau_i^z\right],\label{eq:p4top2_1}\\
    V_i &= \frac{1}{2}\left(\sigma_i^x+\tau_i^x+\mathrm{i}\sigma_i^y\tau_i^z-\mathrm{i}\sigma_i^z\tau_i^y\right),\label{eq:p4top2_2}
\end{align}
where \(\boldsymbol{\sigma}_i\) and \(\boldsymbol{\tau}_i\) denote two decoupled sets of Pauli operators acting on site \(i\). Each copy is Eq.~\eqref{eq:Ising} with $J\to J/2$, and $h\to h/2$. The resulting two Ising models are identical and decoupled, each with $J, h,$ and $h_{\partial}$ halved (leaving $J/h$ and $h_{\partial}/h$ unchanged).

\begin{figure}[t]
    \centering
    \includegraphics[width=0.8\linewidth]{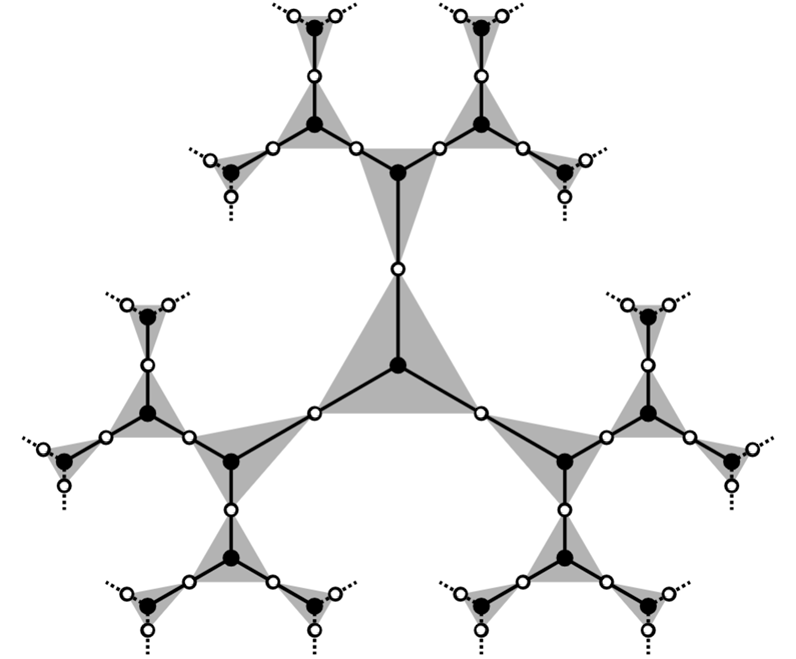}
    \caption{
    Vertex-centered representation of the coordination-$3$ infinite Cayley tree (lines and solid circles) and its dual hyperedge-centered hypertree (triangular shades and hollow circles). Vertices (hollow circles) and $3$-vertex hyperedges (triangular shades) in the dual lattice map to bonds (lines) and vertices (solid circles) in the original lattice, respectively. The dashed lines and light shades indicate continuation of the infinite lattice.  
    }
    \label{fig:BetheLattice}
\end{figure}

The underlying lattice is the coordination-$z$ infinite Cayley tree: the depth-$n$ perfect tree about a vertex or an edge, taken to $n\rightarrow\infty$ (Fig.~\ref{fig:BetheLattice}). The Cayley tree embodies non-amenability. With $\mathcal{V}_n$  denoting the number of vertices of the full depth $n$ tree and $\partial \mathcal{V}_n$ its outer shell,
\begin{equation}
    \lim_{n\rightarrow\infty}\frac{|\partial \mathcal{V}_n|}{|\mathcal{V}_n|} = \frac{(z-2)}{(z-1)} \neq 0. 
    \quad (z>2),
\label{eq:nonamen}
\end{equation}
Thus, the boundary remains an extensive fraction of the system even in the thermodynamic limit.

On a Euclidean lattice, a polarized branch of the ordered phase can be selected by applying a longitudinal field $h_\partial$ at the boundary $\partial$V, 
\begin{equation}
    H_{p\text{-clock}}'=H_{p\text{-clock}}-\Big(\frac{h_\partial}{2}\sum_{i\in\partial\mathrm{V}}U_i+\mathrm{h.c.}\Big),
\label{eq:bdryfield}
\end{equation}
which explicitly breaks the \(\mathbb{Z}_p\) symmetry and biases the clock operators toward the \(+1\) direction. On a non-amenable lattice, however, the boundary field \(h_\partial\) acts on an extensive number of sites, potentially giving rise to ordering with no Euclidean counterpart.

{\it Symmetry-fixing duality.--} For the quantum $p$-clock model, Eq.~\eqref{eq:pclock}, on any finite graph with $h>0$, the Perron--Frobenius theorem guarantees a unique ground state which lies in the unit-eigenvalue sector of $\mathcal{U}$, defined in Eq.~\eqref{eq:zp}. When restricted to this sector, on a loopless graph, such as the coordination-$z$ Cayley tree, the symmetric Hamiltonian, Eq.~\eqref{eq:pclock} without boundary field is unitarily equivalent to 

\begin{equation}
\label{eq:dualmodel}
H_{p\text{-clock}}^*
=
-\frac{J}{2}\sum_{e\in \mathrm{E}}
V_{e}
-\frac{h}{2}\sum_{i\in \mathrm{V}}
\left(
\prod_{e\ni i}U_{e}
\right) + \mathrm{h.c.}\;.
\end{equation}
Since a tree has no cycles, there are no gauge (zero flux) constraints that require oriented products of $V_e$ around each cycle to be the identity.  The dual operators $V_e, U_e$ reside on edges $e$ of $\mathrm{G}(\mathrm{V}, \mathrm{E})$ (therefore vertices of the dual hypergraph; see Fig. \ref{fig:BetheLattice}). The symmetry operator of \eqref{eq:zp} is mapped to identity, $\mathcal{U} \mapsto \mathbb{I}$, therefore realizing the targeted symmetry sector of the original model. In any symmetry sector, the central (bulk) polarization $m_0=|\langle U_0\rangle|$ vanishes identically. However, the bulk long-range correlator (Onsager-Yang) still diagnoses ordering
\begin{equation}
    (m_0)_{\mathrm{sym}}^2:= |\lim_{x\to \infty} \langle U_{(0-x)}U_{(0+x)}^{\dagger}\rangle|,
\end{equation}
which maps onto a string operator in the dual model
\begin{equation}
    \lim_{x\to \infty} \langle U_{(0-x)}U_{(0+x)}^{\dagger}\rangle\mapsto \lim_{x\to \infty} \langle \prod_{e\in \gamma(-x,x)} V_e^{r_e}\rangle,
\end{equation}
where \(\gamma(-x,x)\) denotes the unique path connecting two \textit{bulk} vertices, each at distance \(x\) from the central bond on its two sides, and \(r_e= 1 \) (or $r_e = -1$) if the path runs along (respectively opposite to) the orientation of the  edge $e$. 
Details of the Perron--Frobenius argument and the bond-algebraic duality construction are given in \cite{SM}.

{\it {\rm iDMRG} calculations.--} We use an infinite-system DMRG approach that grows an edge-centered tree, layer by layer, truncating each subtree to $\chi$ states~\cite{White1992}; the physical boundary is kept explicitly. The couplings are parameterized as $K = J/h$ and $B_\partial=h_{\partial}/h$. The original model Eq.~\eqref{eq:pclock} is simulated to give the boundary-biased state, while the dual model Eq.~\eqref{eq:dualmodel} with $B_\partial=0$ is simulated on the dual hypertree to give the symmetric state. We retain $\chi=16$ ($p=2$) and $\chi=18$ ($p=3$) with $B_\partial=10^{-8}$. The iteration is first carried for 1000 cycles and then indefinitely until a convergence threshold of $\delta C_{01}<10^{-8}$ is reached. Each $K$ is grown independently so no sweep history enters. 

We compute the central polarization \(m_0\) and correlator \(C_{01}=\mathrm{Re}\langle U_0U_1^\dagger\rangle\), together with their dual counterparts, \(\big|\langle\prod_{e\in\gamma(-\infty,\infty)}V_e^{r_e}\rangle\big|^{1/2}\) and \(\mathrm{Re}\langle V_{01}\rangle\), respectively. We also evaluate the central-bond entanglement entropy \(S_{01}\) and its dual analogue \(S^*_{abc}\), defined as the von Neumann entropies of the reduced subtrees obtained by cutting the corresponding central bonds. For \(p=2\), we further compute the boundary polarizations \(m_\partial=|\langle U_\partial\rangle|\) and \(x_\partial=|\langle V_\partial\rangle|\), as well as the polarization \(m_{\partial+\alpha}\) at distance \(\alpha\) from the boundary. The boundary-bond entanglement entropy is $S_{\partial} = -(\frac{1+r}{2}) \mathrm{ln}(\frac{1+r}{2}) -(\frac{1-r}{2}) \mathrm{ln}(\frac{1-r}{2})$, with $r=\sqrt{m_\partial^2+x_\partial^2}$.

Bulk observables are converged to a relative accuracy of \(10^{-4}\) between \(\chi=12\) and \(16\) for \(p=2\), and between \(\chi=15\) and \(18\) for \(p=3\), provided \(|K-K_{c_2}|>5\times10^{-4}\). The transition points agree within the scan resolution, \(\Delta K=10^{-4}\). Varying the nonzero boundary bias over \(10^{-8}<B_\partial<10^{8}\) yields indistinguishable bulk results. Boundary observables are evaluated for \(10^{-6}<B_\partial<10^{-5}\) and linearly extrapolated to \(B_\partial\to0^+\), in agreement with those selected by machine-precision bias. Details of the growth and partition procedures are provided in~\cite{SM}.

{\it Boundary-sensitive bulk phase.--} For both the $p=2$ and $p=3$ clock models, we find a robust intermediate phase characterized by the boundary sensitivity of Eq.~\eqref{eq:BSP}: the bulk orders in the boundary-biased state while remaining disordered in the symmetric state.

\begin{figure*}[htbp]
  \centering
  \includegraphics[width=0.97\textwidth]{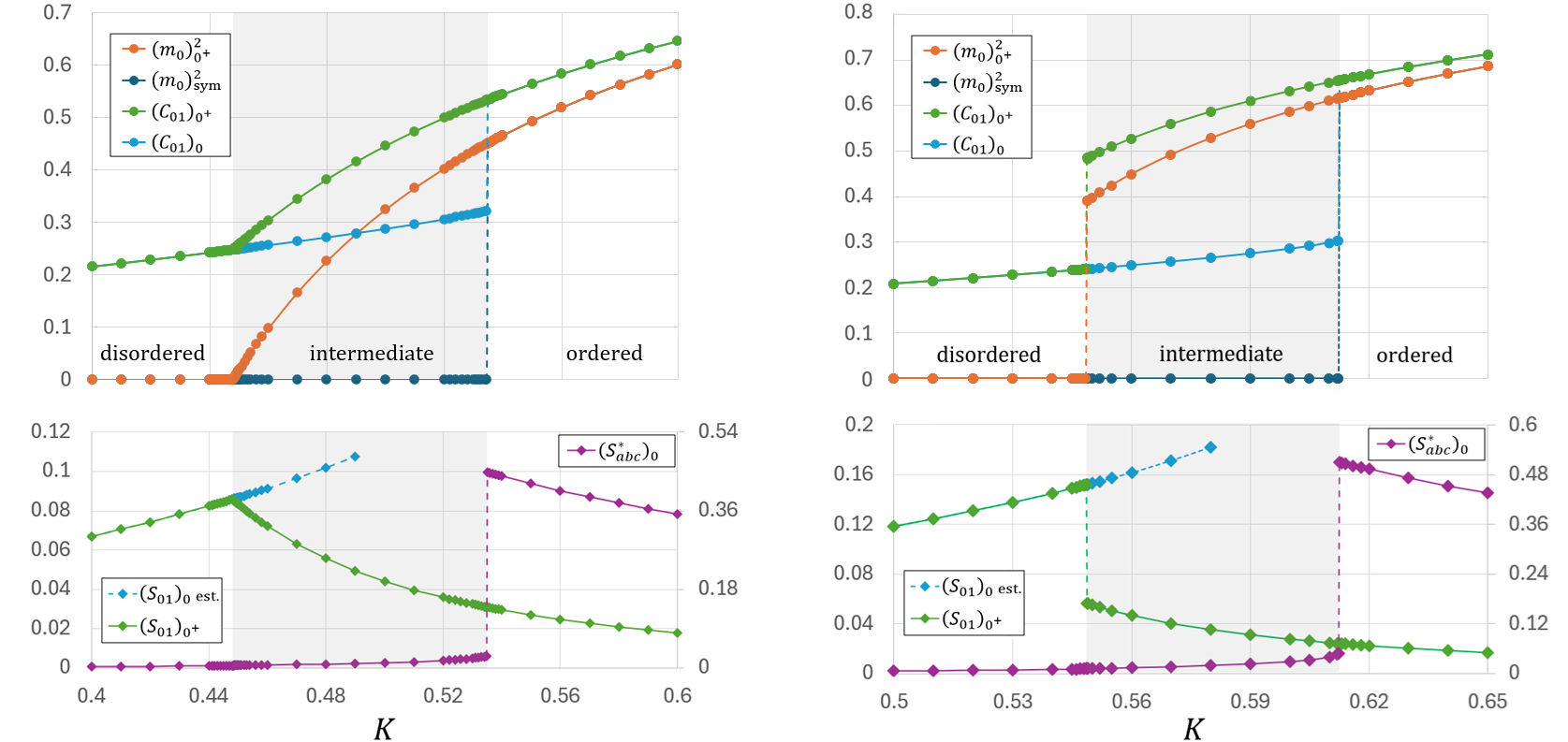}
  \caption{Quantum phase diagram, and various observables, for $p=2$ (left, $\chi=16$) and $p=3$ (right, $\chi=18$). Top: bulk order parameter $m_0^2$ of the boundary-biased (orange) and symmetric (blue) states, and central correlator $C_{01}$ of the biased (green) and symmetric (cyan) states. Bottom: central-bond entropy $S_{01}$ of the biased (green) and symmetric (blue dashed) states\footnote{for small $K$ and $B_{\partial}=0$, an unstable plateau resembling symmetric state can be observed before converging to the biased state. }, and central dual hyperedge entropy $S^*_{abc}$ of the symmetric state (purple, right axis). The top and bottom panels share the same $K$-axis, and the shading marks the intermediate phase. The biased state is singular only at $K_{c1}$ (continuous for $p=2$, discontinuous for $p=3$); the symmetric state only at $K_{c2}$ (discontinuous for both).}
  \label{fig:bulkall}
\end{figure*}

For the \(p=2\) Ising model, Fig.~\ref{fig:bulkall} reveals two distinct transitions. At \(K_{c1}\simeq0.448\), the central polarization of the boundary-biased state turns on continuously, while that of the symmetric state remains zero, marking a continuous transition from the disordered to the intermediate phase. The critical point agrees with previous infinite-tree MPS results~\cite{Nagaj2008,footnote1}. Fitting \(m_0\sim(K-K_{c1})^\beta\) over \(10^{-4}\leq K-K_{c1}\leq10^{-3}\) yields \(\beta\simeq0.50\), consistent with the mean-field value \(\beta=1/2\). At \(K_{c2}\simeq0.535\), the correlator and order parameter of the symmetric state jump to those of the biased state, agreeing within a relative difference of \(10^{-4}\) and signaling a discontinuous transition into the spontaneously ordered phase,  which was not identified in Ref.~\cite{Nagaj2008}. The \(p=3\) clock model exhibits the same three-phase structure, with \(K_{c1}\simeq0.549\) and \(K_{c2}\simeq0.612\), but both transitions are discontinuous. See Fig. \ref{fig:bulkall}.

\begin{figure*}[htbp]
    \centering
    \includegraphics[width=0.97\linewidth]{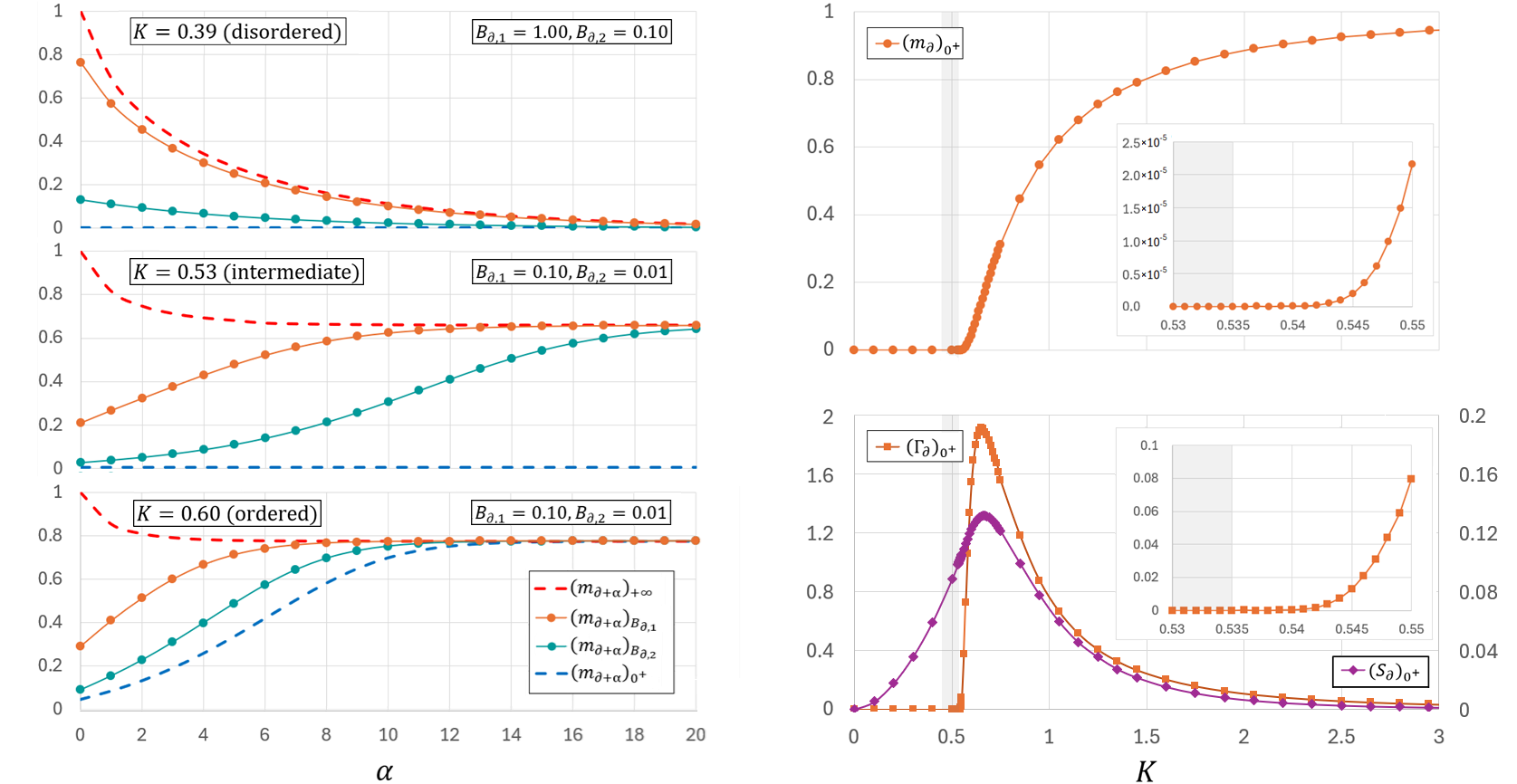}
    \caption{Boundary-to-bulk propagation in the $p=2$ model ($\chi=16$).
Left: layer-resolved magnetization $m_{\partial+\alpha}$ at distance $\alpha$
from the boundary for $K=0.39$, $0.53$, and $0.60$, representing the
disordered, intermediate, and ordered phases, respectively. Dashed curves
show the $B_\partial\to\infty$ and $B_\partial\to0^+$ limits, while solid
curves show two finite boundary fields indicated in each panel.
Right: boundary magnetization $m_\partial$ (top), its derivative
$\Gamma_\partial$, and boundary entanglement
entropy $S_\partial$ (bottom); insets magnify the onset above $K_{c2}$.
The shaded region denotes the intermediate phase.}
    \label{fig:bdryall}
\end{figure*}

{\it Boundary-to-bulk propagation.--}
A defining feature of the intermediate phase is the penetration of boundary perturbations arbitrarily deep into the bulk. Figure~\ref{fig:bdryall} shows the magnetization \(m_{\partial+\alpha}\) at depth \(\alpha\) from the boundary for representative values of \(K\) in the disordered, intermediate, and ordered phases. In the disordered and ordered phases, the boundary-field dependence decays with \(\alpha\), and all profiles converge to a unique bulk limit. In the intermediate phase, by contrast, the boundary response persists to arbitrarily large depths and ultimately approaches a boundary-selected bulk polarization. Consequently, the limits \(B_\partial\to0^+\) and \(\alpha\to\infty\) do not commute
\begin{equation}\hspace*{-0.25cm}
\lim_{\alpha\rightarrow\infty}
\lim_{B_\partial\rightarrow0^+}
\lim_{N\rightarrow\infty}
m_{\partial+\alpha}
\neq
\lim_{B_\partial\rightarrow0^+}
\lim_{\alpha\rightarrow\infty}
\lim_{N\rightarrow\infty}
m_{\partial+\alpha}\;.
\label{eq:nearbdrylimits}
\end{equation}
The thermodynamic bulk therefore retains information about the boundary condition.

Remarkably, the boundary itself remains unpolarized throughout the intermediate phase under perturbative field despite bulk polarization, and polarizes only gradually within the ordered phase, and both the boundary entanglement entropy \(S_\partial\) and the derivative of the boundary magnetization \(\Gamma_\partial=\partial m_\partial/\partial K\) peak only well inside the ordered phase, at \(K\simeq0.68\).



 

{\it Discussion.--}
We have established a form of quantum order unique to non-amenable geometry: boundary-sensitive order. Its defining feature is the disagreement between two thermodynamic states. Below $K_{c1}$, the symmetric and boundary-biased states are both disordered, while above $K_{c2}$ both are ordered (Fig. \ref{fig:bulkall}). Between them, however, the symmetric state remains continuously connected to the disordered phase while an arbitrarily weak boundary bias selects an ordered bulk state, realizing Eq.~\eqref{eq:BSP}. This order is neither spontaneous—the symmetric state remains disordered—nor topological—it is detected by a local order parameter. Rather, it is the quantum counterpart of the two-stage ordering found in classical hyperbolic systems~\cite{Hyperising,CCWU2,WangNussinovOrtiz2025}.

The origin is non-amenability itself. Because \(|\partial\mathcal V_n|\propto|\mathcal V_n|\), an arbitrarily weak field $h_\partial$ acting on the boundary carries an extensive energetic weight. In the conventional ordered and disordered phases, its influence is confined to a finite boundary layer. In the intermediate phase, by contrast, the bulk becomes susceptible to the boundary: an infinitesimal symmetry-breaking bias penetrates arbitrarily deep into the system and selects its bulk state, even while the boundary itself remains unpolarized (Fig.~\ref{fig:bdryall}). A spin deep in the bulk is connected  to a number of boundary sites that is exponential in its depth. In the intermediate phase, this exponential growth notably amplifies the influence of the boundary. Boundary sensitivity  also exposes a fundamental numerical challenge: bulk-only methods discard the extensive boundary, whereas symmetry-unfixed calculations can mistake the boundary-selected branch for spontaneous order. Resolving the intermediate phase therefore requires retaining the boundary while enforcing the symmetry exactly. We achieved this exact symmetry control through a symmetry-fixing duality.

Our construction extends to general \(\mathbb Z_p\) models on arbitrary trees and, subject to the gauge constraints discussed in \cite{SM}, to bipartite hyperbolic lattices. Its extension to continuous symmetries, itinerant systems~\cite{Gotz2024,Lunts2021}, and varying curvature and coordination may reveal broader classes of boundary-sensitive quantum matter. There is also an intriguing inversion of the usual holographic perspective: while hyperbolic holographic constructions encode bulk information at the boundary~\cite{HaPPY2015,Holography1,Holography2}, here an infinitesimal boundary choice propagates inward and selects the thermodynamic state of the bulk. This mechanism should be directly relevant to experimental realizations of hyperbolic matter~\cite{HyperbolicChern2,KollarHyperbolic,LenggenhagerCircuit,ChenHyperbolicMatter,HuangHyperbolicPhotonic,PatinoHyperbolicMechanical}, where the extensive boundary is not an idealization but an unavoidable part of the system. This boundary sensitivity invites comparison with measurement-induced phase transitions, where the scaling of entanglement entropy distinguishes phases \cite{FisherKhemaniNahumVijay}. Here, the distinction lies in whether boundary information survives arbitrarily deep into the bulk as in the purification transition \cite{GullansHuse}.


{\it Acknowledgements.--} G.O. gratefully acknowledges support from the Institute for Advanced Study. We also gratefully acknowledge the hospitality of the 2026 SPARK program at the Donostia International Physics Center (DIPC) and the Aspen Center for Physics, which is supported by National Science Foundation grant PHY-2210452.


\FloatBarrier
\bibliography{ref.bib}
\bibliographystyle{unsrt}

\pagebreak
\clearpage
\onecolumngrid

\setcounter{secnumdepth}{1}
\setcounter{section}{0}
\renewcommand{\thesection}{S\arabic{section}}

\begin{center}
\textbf{\large Supplemental Material: Boundary-Sensitive Quantum Order on Non-Amenable Lattices}
\end{center}

\section{Ground state uniqueness and symmetry sector of quantum clock models}
\label{appB}

In the main text, we invoked the Perron-Frobenius theorem to claim that the ground state of the Ising and generic $p$-clock models on any finite graph is unique and lies in the unit-eigenvalue sector of the global symmetry operator $\mathcal{U}$, Eq.~\eqref{eq:zp}. Here we sketch the proof for the generic case.

We work in the basis
\begin{eqnarray} \hspace*{-0.5cm}
    U_i | \{s_k\} \rangle = \omega^{s_i} | \{s_k\} \rangle , \quad s_k=0,\cdots,p-1 \ , k\in \mathrm{V}.
\end{eqnarray}
Under this basis, the Hamiltonian in Eq.~\eqref{eq:pclock} is real and symmetric: the interaction terms $U_iU_j^\dagger+\mathrm{h.c.}$ are real diagonal, while the rotation-field terms $V_i+V_i^\dagger$ are real off-diagonal and symmetric. With the sign convention used in Eq.~\eqref{eq:pclock}, all off-diagonal matrix elements are non-positive. After adding a sufficiently large constant shift, we may equivalently consider
\begin{equation}
H' = C\mathbb{I}-H ,
\end{equation}
whose matrix elements are all non-negative. The lowest-energy eigenstate of $H$ is then the eigenvector of $H'$ with the largest eigenvalue.

The Perron--Frobenius theorem implies that if the non-negative $H'$ is irreducible, then its largest eigenvalue is nondegenerate and the corresponding eigenvector can be chosen to have strictly positive entries. Equivalently, the ground state of $H$ is unique and can be written as 
\begin{equation}
|\psi_{\rm GS}\rangle = \sum_{\{s_i\}} c_{\{s_i\}} |\{s_i\}\rangle,
\qquad
c_{\{s_i\}}>0 ,
\end{equation}

It remains to determine the symmetry sector of this state. Since the global  $\mathbb{Z}_p$ clock symmetry Eq.~\eqref{eq:zp} commutes with the Hamiltonian Eq.~\eqref{eq:pclock}, the unique ground state must also be an eigenstate of $\mathcal{U}$:
\begin{equation}
\mathcal{U}|\psi_{\rm GS}\rangle = \lambda |\psi_{\rm GS}\rangle .
\end{equation}
In the classical clock basis, each $V_i$ is represented by non-negative permutation matrices that shifts local clock eigenvalues, and therefore $\mathcal{U}=\prod_{i\in\mathrm{V}}V_i$ is non-negative. Because all coefficients of $|\psi_{\rm GS}\rangle$ are strictly positive, $\mathcal{U}|\psi_{\rm GS}\rangle$ also has strictly positive coefficients. Hence $\lambda$ must be real and positive. Since $\mathcal{U}$ is unitary and $\mathcal{U}^p=\mathbb{I}$, its (degenerate) eigenvalues are $1,\omega, ..., \omega^{p-1}$. The only real positive possibility is therefore
\begin{equation}
\lambda=1 .
\end{equation}
Thus the unique ground state lies in the unit-eigenvalue symmetry sector of $\mathcal{U}$.

Finally, we verify the irreducibility condition. For a real symmetric matrix, irreducibility is equivalent to the connectivity of the graph obtained by treating the matrix as an adjacency matrix with all nonzero off-diagonal entries representing edges. In the present case, the off-diagonal part of $H'$ is generated by summing all the single-site clock-shift operators,
\begin{equation}
H'_{\rm offdiag}
=
\frac{h}{2}\sum_{i\in \mathrm{V}}\left(V_i+V_i^\dagger\right)\;\;\;\;(h>0),
\end{equation}
and the vertices of the connectivity graph resembles classical spin configurations. Since $V_i$ acts as single-site clock shift under classical clock basis, two classical configurations are adjacent if and only if they differ by increasing or decreasing the clock value on exactly one site by one step ($\mathrm{mod}\;p$). Starting from any classical configuration, one can reach any other configuration by successively applying such single-site shifts to the vertices where the two configurations differ. Hence the configuration graph is connected, and therefore $H'$ is irreducible. The Perron--Frobenius theorem then establishes that the ground state for any finite graph is unique and has unit-eigenvalue for $\mathcal{U}$.

While degeneracy can emerge the graph becomes infinite,  thermodynamic limit state takes the finite-size sequence and therefore symmetry properties remain unchanged.

\section{Bond-algebraic mapping to specific symmetry sectors on tree graphs}
\label{appC}

In the main text, we introduced a mapping from the $p$-state clock model on the Cayley tree to a dual model acting in the $\mathcal{U}=1$ symmetry sector on its dual hypertree. Although this mapping does not apply to generic graphs without additional constraints, we show here that it is well defined on tree graphs, including infinite Cayley trees.

A graph $\mathrm{G}(\mathrm{V}, \mathrm{E})$ is bipartite if its vertex set $\mathrm{V}$ can be divided into two disjoint subsets $\mathrm{V}_1$ and $\mathrm{V}_2$ such that every edge connects one vertex in $\mathrm{V}_1$ to one vertex in $\mathrm{V}_2$, and no edge connects two vertices within the same subset. Equivalently, a graph is bipartite if and only if it contains no odd cycles. Therefore, every tree graph is bipartite.

For the $p$-clock model in Eq.~\eqref{eq:pclock} on a bipartite graph $\mathrm{G}=(\mathrm{V}_1,\mathrm{V}_2,\mathrm{E})$, we choose an orientation of every edge from $\mathrm{V}_1$ to $\mathrm{V}_2$: For an oriented edge $e=(i,j) \in \mathrm{E}$, with $i\in \mathrm{V}_1$ and $j\in \mathrm{V}_2$, consider the following mapping
\begin{equation}
\label{eq:partialiso1}
U_iU_j^\dagger \mapsto V_{e}, 
\qquad e=(i,j),
\end{equation}
and 
\begin{equation}
\label{eq:partialiso2}
\begin{aligned}
V_i
&\mapsto
\omega^{t_i}\prod_{e\ni i}U_{e}^\dagger,
\qquad i\in \mathrm{V}_1, \qquad t_i=0,1\cdots, p-1,\\
V_j
&\mapsto
\omega^{t_j}\prod_{e\ni j}U_{e},
\qquad j\in \mathrm{V}_2, \qquad t_j=0,1\cdots, p-1 ,
\end{aligned}
\end{equation}
with $\omega=e^{2\pi\mathrm{i}/p}$. Here the dual degrees of freedom live on edges rather than vertices. It is straightforward to verify that the bond algebras Eqs.\eqref{eq:clockalgebra1}\eqref{eq:clockalgebra2}\eqref{eq:clockalgebra3} are preserved under this mapping.

If the graph contains cycles, the edge variables are subject to additional gauge constraints, since the product of edge variables along an oriented cycle $\mathcal{C}$ telescopes: 
\begin{equation}
\label{eq:gaugeconstraint}
\prod_{e\in \mathcal C} V_e^{r_e}
\mapsfrom
\prod_{\ell}
U_{v_\ell}U_{v_{\ell+1}}^\dagger
\equiv
\mathbb{I},
\end{equation}
where $r_e=+1$ if the traversal of $\mathcal C$ agrees with the chosen orientation of $e$, and $r_e=-1$ otherwise. For tree graphs, no such constraint arises, and edge variables are therefore independent. 

Combining Eqs.~\eqref{eq:partialiso1} and \eqref{eq:partialiso2}, the $p$-clock Hamiltonian in Eq.~\eqref{eq:pclock} is mapped to
\begin{equation}
\label{eq:partialiso3}
H_{p\text{-clock}}
\mapsto
-\frac{J}{2}\sum_{e\in \mathrm{E}}(V_{e}+V_{e}^\dagger)
-\frac{h}{2}\sum_{i\in \mathrm{V_1}}
\left(\omega^{-t_{i}}\prod_{e\ni i}U_{e}+\omega^{t_{i}}\prod_{e\ni i}U_{e}^\dagger
\right)-\frac{h}{2} \sum_{j\in \mathrm{V_2}}\left(\omega^{t_{j}}\prod_{e\ni j}U_{e}+\omega^{-t_{j}}\prod_{e\ni j}U_{e}^\dagger
\right)
\end{equation}
with the global $\mathbb{Z}_p$ symmetry operator mapped as:
\begin{equation}
\begin{aligned}
\mathcal{U}
&=
\prod_{i\in \mathrm{V}}V_i  \\
&\mapsto
\left(\prod_{i\in \mathrm{V}_1}
\omega^{t_i}\prod_{e\ni i}U_{e}^\dagger\right)
\left(\prod_{j\in \mathrm{V}_2}
\omega^{t_j}\prod_{e\ni j}U_{e}\right)  \\
&=
\left(\prod_{e\in \mathrm{E}}
U_{e}U_{e}^\dagger\right)\left(\prod_{i\in \mathrm{V}}\omega^{t_i}\right)
=
\omega^{t}\mathbb{I},\qquad t:=(\sum_{i\in\mathrm{V}}t_i) \;\mathrm{mod} \; p. 
\end{aligned}
\end{equation}
In the second line, we used the fact that every edge of a bipartite graph has one endpoint in $\mathrm{V}_1$ and the other endpoint in $\mathrm{V}_2$. Since the global $\mathbb{Z}_p$ symmetry operator is mapped as a constant $\omega^t$, the bond algebra mapping realizes this specific symmetry sector of the original model Eq.~\eqref{eq:pclock}. 

By setting $t_i\equiv0$, one obtains $\mathcal{U}=1$, which is the proposed dual Hamiltonian Eq.~\eqref{eq:dualmodel}. The bulk translation symmetry of the infinite Cayley tree was restored despite the bipartition of vertices and the orientation of edges in Eqs.~\eqref{eq:partialiso1} and \eqref{eq:partialiso2}, since under this setting exchanging $\mathrm{V}_1$ and $\mathrm{V}_2$ produces the Hermitian conjugate, which is also present in the Hamiltonian. 

The bond-algebra mapping should be better understood as a partial-isometry. This can be demonstrated by considering the symmetry sector projectors on the original Hilbert space: 

\begin{equation}
    P_a = \frac{1}{p}\sum_{n=0}^{p-1}\omega^{-an}\mathcal{U}^n,
\end{equation}
where $a=0,1,2,\ldots,p-1 $, and the corresponding symmetry charge is $\omega^a$. The dimensionality of each symmetry sector is: 

\begin{equation}
\begin{aligned}
    \mathrm{dim}(\mathcal{H}_a) & = \mathrm{Tr}(P_a) = \frac{1}{p}\mathrm{dim}(\mathcal{H}_\mathrm{full}) \\
    & = p^{|\mathrm{V}|-1}
\end{aligned}
\end{equation}
since $\mathbb{I}=\mathcal{U}^0$ is the only term of nonzero trace, and the full Hilbert space (vertex space) has dimensionality $p^{|\mathrm{V}|}$, with $|\mathrm{V}|$ the number of vertices in $\mathrm{G}(\mathrm{V}, \mathrm{E})$. 

The symmetry sector projectors are mapped as: 

\begin{equation}
    P_a \mapsto 
\begin{cases}
  \mathbb{I} & ,a =t, \\
  0   & ,\text{otherwise}. 
\end{cases}
\end{equation}
The dimensionality of the edge space satisfies
\begin{equation}
    \mathrm{dim}(\mathcal{H}_{\mathrm{edges}})=p^{|\mathrm{E}|} = \mathrm{dim}(\mathcal{H}_{0})
\end{equation}
since $|\mathrm{E}|=|\mathrm{V}|-1$ for tree graphs. Thus, the mapping Eqs.~\eqref{eq:partialiso1} and \eqref{eq:partialiso2} reproduces exactly the projection of the $p$-clock model to its $\omega^{t}$ symmetry sector. 

When $\mathrm{G}=(\mathrm{V}, \mathrm{E})$ is an open one-dimensional chain, the mapping is reduced to the usual self-duality mapping of the one-dimensional $p$-clock model.

\section{Infinite-DMRG scheme for the Cayley tree and its dual hypertree}
\label{appA}

Here we describe the infinite-DMRG schemes used in the main text for the $p$-clock model in Eq.~\eqref{eq:pclock} on the coordination-3 infinite tree, as well as for its $\mathcal{U}=1$ partial-isometry representation in Eq.~\eqref{eq:dualmodel} on the dual hypertree. We regard the finite system as two perfect binary trees of equal depth whose roots are connected by a central bond (edge-centered). The coordination-3 infinite Cayley tree is obtained in the limit where the depth of the two subtrees is taken to infinity. The infinite-DMRG scheme approaches this limit by iteratively increasing the subtree depth.

\begin{figure}[!htpb]
\hspace*{-0.5cm}
\includegraphics[width=0.5\textwidth]{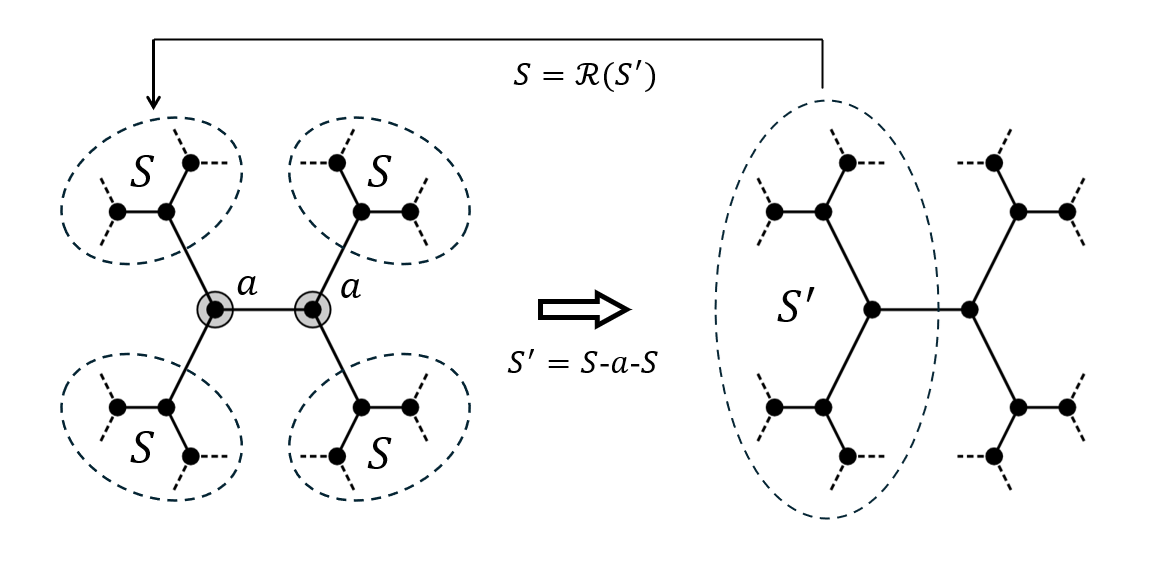}%
\caption{
Infinite-DMRG scheme for the coordination-3 Cayley tree. Black dots denote vertex spin degrees of freedom, while solid and dashed lines denote graph edges. Left: the superblock is assembled from four renormalized subtree blocks $S$ and two newly added vertices $a$. Right: the enlarged subtree $S'=S-a-S$ is truncated to define the new renormalized subtree block $S=\mathcal{R}(S')$ for the next iteration.
}
\label{fig:BetheIDMRG}
\end{figure}

For the original clock model Eq.~\eqref{eq:pclock}, the local degrees of freedom live on the vertices of the infinite tree. As shown in Fig.~\ref{fig:BetheIDMRG}, each iteration uses four identical renormalized subtree blocks $S$ and two newly added vertices $a$ to assemble the superblock. The enlarged subtree
\begin{equation}
\boxed{S'} = \boxed{S}\raisebox{1pt}{---\boxed{a}---}\boxed{S}
\end{equation}
is embedded in the superblock environment, and its reduced density matrix is obtained from the superblock ground state. The new renormalized subtree block is then defined by retaining the $\chi$ dominant eigenstates of its reduced density matrix:
\begin{equation}
S = \mathcal{R}(S') .
\end{equation}
Each iteration increases the depth of the subtree block $S$ by one.

\begin{figure}[!htpb]
\hspace*{-0.5cm}
\includegraphics[width=0.5\textwidth]{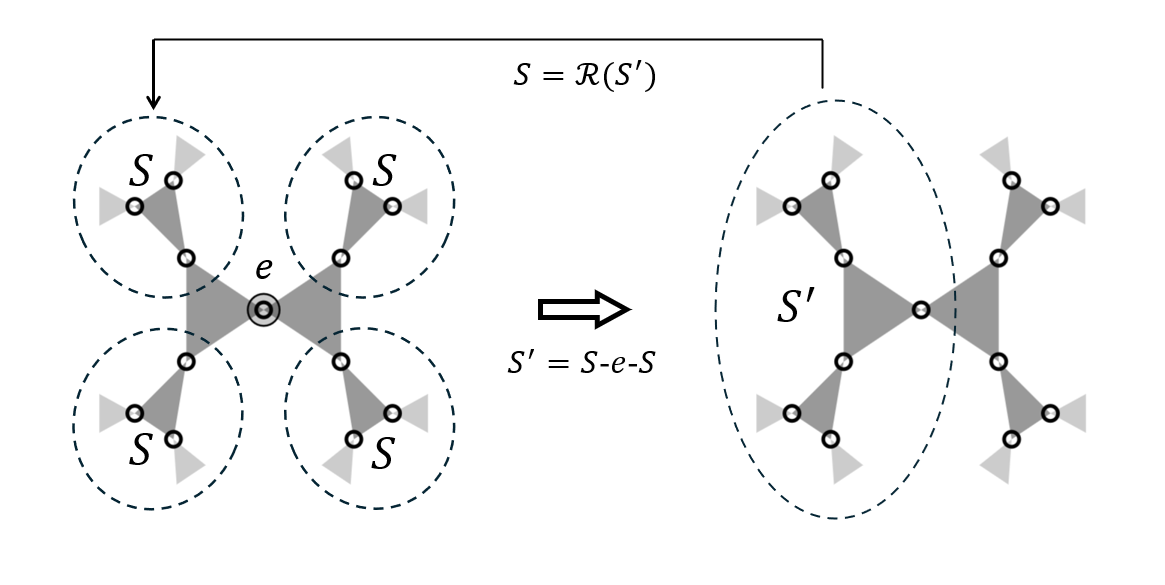}%
\caption{
Infinite-DMRG scheme for the dual of the coordination-3 Cayley tree. Open circles denote edge degrees of freedom of the original lattice, while shaded triangles denote hyperedges of the dual hypertree. Left: the superblock is assembled from four renormalized subtree blocks $S$ and one newly added edge degree of freedom $e$. Right: the enlarged subtree $S'=S-e-S$ is truncated to define the new renormalized subtree block $S=\mathcal{R}(S')$ for the next iteration.
}
\label{fig:DualBetheIDMRG}
\end{figure}

For the dual model Eq.~\eqref{eq:dualmodel}, the degrees of freedom live on the edges of the original infinite tree, and therefore on the vertices of the dual hypertree. The interaction terms of the dual Hamiltonian are supported on hyperedges associated with the original vertices. The corresponding infinite-DMRG step is shown in Fig.~\ref{fig:DualBetheIDMRG}. In analogy with the original lattice update, the superblock is assembled from four identical renormalized subtree blocks $S$ and one newly added edge degree of freedom $e$. The enlarged subtree
\begin{equation}
\boxed{S'}=\boxed{S}\raisebox{1pt}{---\boxed{e}---}\boxed{S}
\end{equation}
is embedded in the superblock environment. After obtaining the superblock ground state, the renormalized subtree block is defined by retaining $\chi$ dominant eigenstates of the reduced density matrix of $S'$. Each iteration again increases the subtree depth by one.

These schemes above are direct generalizations of White's infinite-system DMRG construction for one-dimensional chains. They can also be extended to Cayley trees with higher coordination number in principle. However, the computational cost increases rapidly with the coordination number $z$ since the full superblock contains multiple renormalized subtrees. For coordination number $z\geq 2$, the superblock Hilbert-space dimension scales as
\begin{equation}
\dim \mathcal{H}_{\mathrm{SB}}
= \chi^{2z-2} p^2 ,
\end{equation}
where $p$ is the local clock dimension . For the dual update in Fig.~\ref{fig:DualBetheIDMRG}, the corresponding scaling is
\begin{equation}
\dim \mathcal{H}_{\mathrm{SB}}^{*}
= \chi^{2z-2} p .
\end{equation}
Thus, calculation at larger $z$ is increasingly costly.

\end{document}